\documentclass[fleqn,twoside]{article}
\usepackage[headings]{espcrc2}
\usepackage{graphicx}
\usepackage[figuresright]{rotating}

\newcommand{\AmS}{{\protect\the\textfont2
  A\kern-.1667em\lower.5ex\hbox{M}\kern-.125emS}}
 \title{Search for the existence of circulating currents in high-$T_c$ superconductors\\
 using the polarized neutron scattering technique}

\author{Y.~Sidis\address[LLB]{Laboratoire L\'{e}on Brillouin, CEA-CNRS, CEA-Saclay, 91191 Gif sur Yvette, France\\},
B. Fauqu\'e\addressmark[LLB],
V.~Aji\address[Riverside]{Department of Physics,University of California,Riverside, CA 92521, USA}
 and P.~Bourges\addressmark[LLB]\thanks{To whom correspondence should be addressed;
 E-mail: bourges@llb.saclay.cea.fr}
}
\runtitle{2-column format camera-ready paper in \LaTeX}
\runauthor{Y. Sidis}

\begin{document}
\begin{abstract}
We review experimental attempts using polarized neutron scattering technique to reveal 
the existence in high temperature superconductors of a long-range ordered state 
characterized by the spontaneous appearance of current loops. We draw a particular 
attention to our recent results (B. Fauqu\'e {\it  et al.}, {\it Phys. Rev. Lett.} {\bf 96}, 197001 (2006)) that, up to now, 
can be explained only by the theory of circulating current proposed by C.M. Varma.
\end{abstract}
\maketitle





Since the discovery of high-$T_c$ copper oxide superconductors, a large variety of 
theoretical models has been proposed to capture the intrinsic microscopic nature of $\rm CuO_2$ 
planes, where superconductivity develops. Several models postulate the existence of circulating 
currents (or current loops) in the $\rm CuO_2$ planes that could be responsible for the exotic 
electronic properties of these materials, in particular the existence of the mysterious 
"pseudo-gap" phase. In underdoped high-$T_c$ copper oxide superconductors, this phase is 
notable from its departure from the behavior of conventional metal 
\cite{workshop,revue,Timusk99,Tallon01,Alloul89,Ito93,Loram94} and it has been proposed that 
it could correspond to a long-range ordered phase associated with a new state of matter, 
that could either co-exist or compete with superconductivity. 
\begin{figure}[htb]
\includegraphics[width=5.5cm,angle=270]{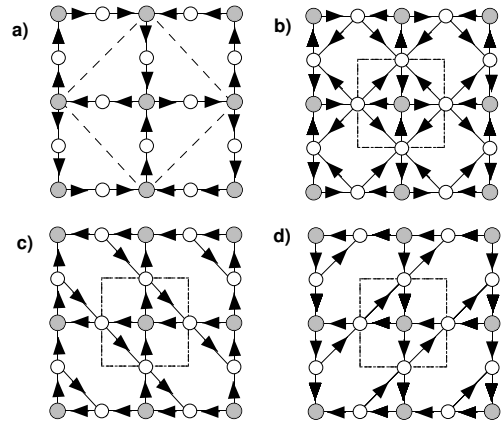}
\vspace{-0.75cm}
\caption{Orbital currents in $\rm CuO_2$ plane (coppers: filled circles, oxygens: opened circles ):
a) DDW state, b) CC-phase $\Theta_I$, c-d) two domains of CC-phase $\Theta_{II}$. 
The dashed lines indicate the size of the related Brillouin zone. }
\vspace{-0.75cm}
\label{fig-CCpatterns}
\end{figure}

In the staggered flux phase model \cite{Hsu91} or {\it d-wave} charge density wave (DDW) 
model \cite{Chakravarty00,Nayak00,Chakravarty01,Chakravarty01-neutron}, $\rm CuO_2$ planes 
are characterized by the appearance of staggered circulating currents flowing into square 
copper plaquettes (Fig.~\ref{fig-CCpatterns}.a).
In contrast to these models, where both translation invariance and time reversal symmetry are 
broken, the time reversal symmetry is broken in the circulating current (CC) phase proposed by 
C.M. Varma, but translation invariance is preserved \cite{Varma97,Varma99,Simon02,Varma06}. 
The order parameter of CC-phase is made of a set of staggered circulating currents  embedded 
in a given Cu square plaquette. Each current flows along a triangular path involving a copper 
site and two of its nearest neighbor oxygen sites (Fig.~\ref{fig-CCpatterns}.b-d).
Two distinct currents patterns have been proposed: the $\Theta_I$-phase made of 4 triangular 
loops (Fig.~\ref{fig-CCpatterns}.b) or the $\Theta_{II}$-phase made of 2 triangular loops 
(Fig.~\ref{fig-CCpatterns}.c) forming a "butterfly" centered on  
Cu site (Fig.~\ref{fig-CCpatterns}.c-d)

\begin{figure}[t]
\vspace{0cm}
\includegraphics[width=7cm,angle=0]{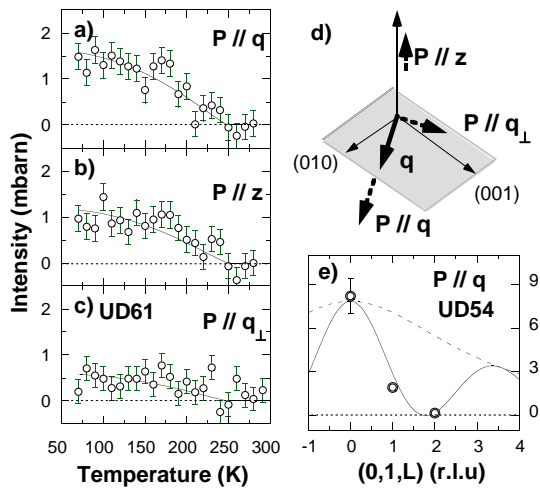}
\vspace{-0.75cm}
\caption{Temperature dependencies of the normalized magnetic intensity, 
measured in the SF channel at 
$\bf{q}$=(0,1,1) in sample UD61:
a) for $\bf{P}$//$\bf{\hat{q}}$,
b) for $\bf{P}$//$\bf{\hat {z}}$,
c) for $\bf{P}$//$\bf{\hat{q}_{\perp}}$ 
d) Sketch of the scattering plane showing the three polarization directions discussed 
here.
e) Sample UD54: L-dependence of the normalized magnetic intensity measured at T=70 K 
and $\bf{q}$=(0,1,1) as compared to $\frac{cos^2(\pi L d/c)}{q^2}$ (solid line) 
and  $\frac{1}{q^2}$ (dashed line).
} 
\vspace{-0.75cm}
\label{figUD61}
\end{figure}

Whatever its origin, the array of circulating currents in the $\rm CuO_2$ planes generates 
a magnetic field distribution: for the staggered flux phase or the DDW phase, one may speak 
about antiferromagnetic (AF) orbital magnetism (with a propagation wave vector at 
{\bf q}=$(\pi,\pi)$ and {\bf q}=0 AF-orbital magnetism in the case of CC-phases.
Neutron scattering technique can be used to probe the existence of circulating currents in 
high-$T_c$ copper oxide superconductors. Neutron spin does indeed interact  with any 
internal magnetic field. The magnetic intensity for scattering neutron is
written\cite{squires}:
\begin{equation}
(\frac{d \sigma}{d \Omega})_{i \rightarrow f} \propto \vert < f \vert \sigma.{\bf B}({\bf q}) \vert i > \vert ^2
\label{eq0}
\end{equation}
where $\sigma$ is the neutron spin and  $\vert i >$ and $\vert f >$ are the initial and final 
states of the neutron. ${\bf B}({\bf q})$ describes the Fourier transform of the magnetic field 
distribution at the wave vector {\bf q}. When the internal magnetic field originates from the 
circulating currents \cite{Hsu91,Chakravarty01-neutron},  ${\bf B}({\bf q})$ can be expressed 
in terms of the Fourier transform of the current density ${\bf j}({\bf q})$, {\it i.e.} 
${\bf B}({\bf q}) \propto \frac{ {\bf j}({\bf q}) \times {\bf q}}{q^2}$. 
The neutron scattering cross section for polarized neutron then reads\cite{squires}:
\begin{eqnarray}
(\frac{d \sigma}{d \Omega})_{i \rightarrow f} &\propto& \vert < f \vert {\bf \sigma}  \frac { {\bf j}({\bf q}) \times {\bf q} }
{q^2} \vert i >\vert ^2 
\label{eq4}
\end{eqnarray}
The neutron scattering intensity in the spin  flip (SF) channel, $I$, has to fulfill 
a polarization selection rule equivalent of the one for spin  moments which 
in absence of chirality can be written:
\begin{equation}
I_{{\bf P}//{\bf \hat{q}}}= I_{{\bf P}//{\bf \hat{z}}}+I_{{\bf P}//{\bf \hat{q}}_{\perp}}
\label{master-eq1}
\end{equation}
where  the unitary vectors ${\bf \hat{q}}$ and ${\bf \hat{q}}_{\perp}$ are respectively parallel 
and perpendicular to  wave vector ${\bf q}$ in the scattering plane, and ${\bf \hat{z}}$ 
is perpendicular the scattering plane (Fig.~\ref{figUD61}.d). 
For unpolarized neutron and taking into account the charge conservation constraint, 
the cross-section takes the simple form:
\begin{equation}
\frac{d \sigma}{d \Omega} \propto \frac{<j({\bf q})>^2}{q^2}  
\label{eq1}
\end{equation}


The description of the current density could be quite complicated due to the nonzero size of 
the Wannier function, the mixing of different orbital, and many body effects when the current 
becomes substantial \cite{Hsu91,Chakravarty01-neutron}. For sake of  simplicity, it can be 
useful to associate an effective orbital magnetic moment to a given current loop. The moment 
is located at the center of the current loop and is aligned perpendicular to the $\rm CuO_2$ 
planes, owing to the planar confinement of the currents. Therefore, in contrast to a spin moment on Cu 
site, the effective orbital moment should be much more spread in real space. As a consequence, 
one may expect circulating currents to give rise to a neutron scattered magnetic intensity 
decreasing faster than the squared Cu magnetic form factor, $F^2_{Cu}(q)$, at large wave 
vector in agreement with Eq.~\ref{eq1}. 


Let us consider, at first, the case of AF orbital magnetism. There were several observations of 
an AF ordering in the underdoped regime of
superconducting $\rm YBa_2Cu_3O_{6+x}$ \cite{Sidis01,Mook01,Mook02}. When observed,  this phase 
usually develops well above $\rm T_c$ (close to room temperature) and displays an ordered 
moment of about 0.05-0.02 $\rm \mu_B$. The polarization neutron analysis reveals that the 
AF phase is dominated by moments aligned in the $\rm CuO_2$ plane, as in the insulating  
AF parent compound. Considering, on the one hand, that other high quality single crystals 
with a similar doping level do not show a similar order \cite{Stock02} and, on the other 
hand, that impurity substitution out of the $\rm CuO_2$ planes (in the CuO chains) could induce 
an AF order at $\sim$ 300K in optimally doped samples ($\rm T_c$=93 K) \cite{Hodges02}, 
it is likely that the observed AF order is not a generic property of the underdoped state 
and may be either remnant of the AF insulating state or induced by impurities or defects 
in the CuO chains.

Beyond this AF spin order, Mook {\it et al.} \cite{Mook02,Mook04} reported polarized neutron 
experiments suggesting the existence of a weak AF quasi-2D order, potentially characterized 
by: (i) magnetic moments perpendicular to the $\rm CuO_2$ plane, (ii ) a decrease of the 
neutron scattering intensity at large wave vector larger that $F^2_{Cu}(q)$. These experiments 
can be viewed as providing evidence in favor of the DDW model. However, it is worth pointing 
out that the estimated magnitude of the ordered moment is $\sim$ 0.0025$\rm \mu_B$, i.e close 
to the experimental threshold of detection \cite{Mook04}. Furthermore, the small number of 
studied Bragg reflections was not sufficient to get solid conclusions concerning the evolution 
of the magnetic signal at large wave vector. Finally, the lack of reproducibility of the 
neutron scattering data severely questions their relevance for cuprates in general. 
In addition, the study of the charge excitation spectrum by angle resolved photo-emission 
do not show any indication of a doubling of the unit cell, resulting from breaking of 
the translation symmetry expected for the DDW phase. Therefore, there are no compelling 
evidences of circulating currents breaking translation invariance and leading to an 
orbital AF state.


Concerning the two possibles CC phases preserving translation symmetry
proposed by C.M. Varma \cite{Varma97,Varma99,Simon02,Varma06}, the $\Theta_{I}$-phase 
was not detected by earlier polarized elastic neutron scattering 
experiments \cite{Lee99,Bourges98}. The $\Theta_{II}$-phase has been later on
proposed \cite{Simon02} to account for an angle resolved photoemission 
experiment \cite{kaminski}. Using circularly polarized photons, 
A. Kaminski {\it et al} \cite{kaminski} reported a dichroic signal in the 
$\rm Bi_2 Sr_2 Ca Cu_2 O_{8+\delta}$ system indicating a time reversal breaking symmetry 
in the pseudo-gap state. 

Recently, this $\Theta_{II}$-phase 
has been searched with polarized neutron scattering using uniaxial polarization 
analysis \cite{Fauque06}. In contrast to the earlier inconclusive attempts, 
experimental evidences in favor of the existence of the CC $\Theta_{II}$-phase 
were found \cite{Fauque06}: a magnetic order, hidden in the pseudo-gap regime of 
high-$\rm T_c$ superconductors, has been observed. The study performed by 
Fauqu\'e {\it et al.} \cite{Fauque06} covers a large part of the phase diagram 
of high-$\rm T_c$ cuprates and show a remarkable reproducibility of the neutron
results, that was missing in the previous studies devoted to DDW phase. 
Five different samples of different origin were indeed studied: 4 $\rm YBa_2Cu_3O_{6+x}$ 
samples in the underdoped (UD :$\rm T_c < T_{c,max}$) regime and 
one $\rm Y_{0.85}Ca_{0.15}Ba_2Cu_3O_{6+x}$ in the overdoped regime ($\rm T_c > T_{c,max}$). 
Hereafter, the samples are labeled as: UD54 (x=0.55, $\rm T_c$=54 K),
UD61 (x=0.6 - $\rm T_c$=61 K), UD64 (x=0.6 - $\rm T_c$=64 K), UD68 (x=0.75 - $\rm T_c$=68 K),
OD75 (x=1 - $\rm T_c$=75 K). Polarized neutron data have been obtained 
on the triple axis spectrometer 4F1 at the Orph\'ee reactor (Saclay) 
(see Ref. \cite{Fauque06} for the description of the experimental set up).

\begin{figure}[t]
\vspace{0cm}
\includegraphics[width=5cm,angle=270]{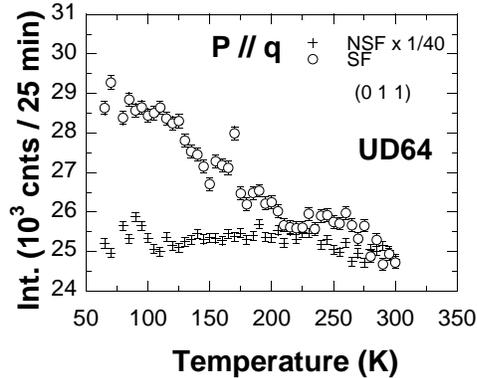}
\vspace{-0.75cm}
\caption{ Temperature dependencies of the raw SF  ($\circ$) and NSF  ($+$) neutron intensity measured 
at $\bf{q}$=(0,1,1) in sample UD64 (from \cite{Fauque06}).}
\vspace{-0.75cm}
\label{figUD64}
\end{figure}

In contrast to the DDW phase or the flux phase, the CC-phase does not break translation 
invariance and the magnetic scattering intensity appears at the same wave vector  as the 
nuclear scattering intensity. For the CC-phase $\Theta_{II}$, both intensities are 
superimposed since interference terms with the nuclear structure factor cancel out 
due the existence of magnetic domains. The use of polarized neutron is then essential to 
identify the magnetic scattering generated by the circulating currents.
Most of the data of Ref.~\cite{Fauque06} were obtained in a scattering plane  
(010)/(001)(Fig.~\ref{figUD61}). In order to evidence small magnetic intensity, 
measurements have to be performed on the weakest nuclear Bragg peaks where the
magnetic scattering is expected: the Bragg peak $\bf{q}$=(0,1,1) offers the best compromise.
So far, only the polarized neutron data in the normal state are reliable. The effects specific 
to the superconducting state such as the evolution of the magnetic response associated 
with the circulating current through $\rm T_c$ remains to be settled carefully.

\begin{figure}[t]
\vspace{0cm}
\includegraphics[width=6.5 cm,angle=270]{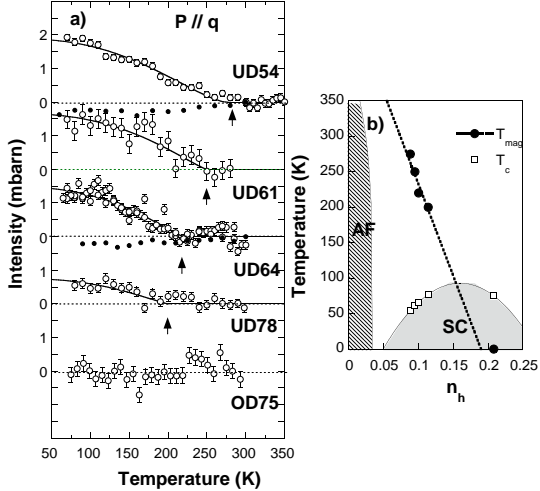}
\vspace{-0.75cm}
\caption{a) Temperature dependencies of the normalized magnetic intensity (see text), 
measured in the SF channel for 5 different samples: at ${\bf q}=(0,1,1)$ ($\circ$) and 
at ${\bf q}=(0,0,2)$ ($\bullet$). b) Phase diagram as a function of hole doping \cite{Fauque06}. 
} 
\label{fig-phase-diagram}
\vspace{-0.75cm}
\end{figure}

Figure \ref{figUD64}.a shows the raw neutron intensity measured at 
$\bf{q}$=(0,1,1) for SF and NSF channel with ${\bf P}//{\bf \hat{q}}$ (sample UD64), 
for which the magnetic scattering is entirely spin-flip and given by Eq.~\ref{eq1}. 
When cooling down from room temperature, the NSF and SF intensities display the same 
evolution within error bars, until  T$<$T$_{mag}$, where the SF intensity increases 
noticeably, whereas the SF intensity remains flat. This behavior signals the presence 
of a spontaneous magnetic order below T$_{mag}$. The normalized magnetic intensity 
$I_{mag}$ (see Ref.~\cite{Fauque06}) is $\sim$ 1 mbarn (Fig.\ref{fig-phase-diagram}.a), 
{\it i.e.} $\sim$ 10$^{-4}$ of the strongest Bragg peaks: it is therefore impossible 
to detect it with unpolarized neutron diffraction. 

The unusual magnetic order in a temperature and doping range that cover the range where 
the pseudo-gap state is observed in  $\rm (Y,Ca)Ba_2Cu_3O_{6+x}$. For the 4 underdoped 
samples, the  full  normalized magnetic intensity  systematically increases  below a 
certain temperature T$_{mag}$ whereas no magnetic signal is 
observed in the overdoped sample (OD75) (Fig. \ref{fig-phase-diagram}).
The deduced $T_{mag}$, defined as the change of slope in the normalized 
intensity  $I_{mag}$, decreases with increasing doping and  matches quite well the pseudo-gap 
temperature, T$^{*}$, of the resistivity data in $\rm YBa_2Cu_3O_{6+x}$ \cite{Ito93,Fauque06}. 
Likewise, $T_{mag}$ decreases almost linearly and extrapolates to 0 K close to the hole doping 
$n_h \simeq$ 0.19, the end point of the pseudo-gap phase \cite{Tallon01} 
(Fig. \ref{fig-phase-diagram}.b).

To proof that the signal reported here is magnetic, it should follow the polarization 
selection rule given by Eq.\ref{master-eq1}. The polarization analysis, shown in 
Fig.\ref{figUD61}.a-c, indicates that the observed signal follows the expected 
selection rule. That unambiguously demonstrates the magnetic origin of the 
low temperature signal. Moreover, the observed magnetic peak  at $\bf{q}$=(0,1,1) is 
resolution limited along the (001) direction, showing that the magnetic  order is 
characterized by long range 3D correlations (at least at T=75 K)\cite{Fauque06} . 
The polarized neutron data  of Ref.~\cite{Fauque06} therefore provide compelling evidences 
of the existence of a magnetic order hidden in the pseudo-gap state of high-$\rm T_c$ 
cuprate superconductors. But can this magnetic order be ascribed to the $\Theta_{II}$-phase ?


For comparison of the neutron scattering data with what one could expect from the 
$\Theta_{II}$-phase, one can express the magnetic intensity  either in term of orbital 
magnetic moments (Ref.~\cite{Fauque06}) or in term of the Fourier transform of the 
current density: both approaches being equivalent at a qualitative level.
Here, we propose to consider the second approach. Since the $\Theta_{II}$-phase can 
be viewed as an array of {butterfly} triangular circulating currents  centered on 
each Cu site  (Fig~\ref{fig-CCpatterns}.c), the Fourier transform of the current 
distribution takes the general form:
\begin{eqnarray}
j({\bf q}) &\propto& \sum_{{\bf G}} \delta_{{\bf q},{\bf G}} \beta(q) \nonumber \\
 &\times& (cos(\pi \frac{H}{2}) A_{10}({\bf q}) {\bf \hat{e}}_x  \nonumber \\
&-& cos(\pi \frac{K}{2}) A_{01}({\bf q}) {\bf \hat{e}}_y \nonumber \\
&+& cos(\pi \frac{H-K}{2}) A_{1-1}({\bf q}) ({\bf \hat{e}}_y -{\bf \hat{e}}_x)  ) 
\label{eq3}
\end{eqnarray}
with ${\bf \hat{e}}_i$ the unitary vector in Cartesian coordinates and ${\bf G}$ vectors 
of the Bravais lattice. According to this expression, the neutron intensity could be 
different from zero {\it only} for wave vectors along ${\bf q}$=\{(0,K,L),(H,0,L),(H,H,L)\}.  
Furthermore, since the current loops are staggered in each Cu plaquette, there is no 
uniform magnetic field in the $\rm CuO_2$ planes, this implies that the neutron scattering 
intensity vanishes for ${\bf q}$=(0,0,L). The magnetic signal is indeed experimentally 
observed at ${\bf q}$=(0,1,L)  (Fig.~\ref{figUD64}-\ref{figUD61}) and vanishes at 
${\bf q}$=(0,0,2) (Fig.~\ref{fig-phase-diagram}.a) ). 

When the circulating currents are modeled with a set of thin wires along which flow a 
current with a given intensity, one obtains for the  $A_{ij}$ terms in Eq. \ref{eq3}:
$A_{ij}({\bf q})\propto sin(\pi (iH+jK) / 2)/(iH+jK)$. 
The terms $A_{ij}$ become either 1 or 0 at a Bragg position. The term $\beta(q)$ is equal
to 1 for a monolayer system. Since $ \rm YBa_2Cu_3O_{6+x}$ is a bilayer systems, current 
loops in the 2 $\rm CuO_2$ planes could be either in phase ($\psi=0$) or out-of-phase 
($\psi=\pi / 2$), so that $\beta(q) \propto cos(\pi L d/c + \psi)$ ($d$=3.3~\AA~ stands
 for the distance between the planes within  the bilayer.) Finally, for the scattering 
plane (010)/(001), the neutron cross-sections, given by Eqs. ~\ref{eq4}-\ref{eq3}, 
can be written for the circulating current model shown Figs.~\ref{fig-CCpatterns}.c-d as
\begin{eqnarray}
I_{{\bf P}//{\bf \hat{q}}} &\propto& \frac{cos^2(\pi L d/c + \psi)}{q^2}  
 \label{master-eq2}
\\
I_{{\bf P}//{\bf \hat{q}}}&=& I_{{\bf P}//{\bf \hat{z}}} 
 \label{master-eq2bis}
\\
I_{{\bf P}//{\bf \hat{q}}_{\perp}}&=& 0 
\label{master-eq3}
\end{eqnarray}

The L dependence of the magnetic Bragg intensity measured in the SF channel 
for ${\bf P}// {\bf \hat{q}}$ and $\bf{q}$=(0,1,L) is in  agreement with 
Eq.~\ref{master-eq2} with $\psi=0$ (Fig.~\ref{figUD61}.e). One may speculate that 
this in-phase configuration could favor the tunneling of charge carriers along the 
c axis. However, in Fig~\ref{figUD61}.a-c, the magnetic signal for $I_{{\bf P}//{\bf \hat{q}}}$ 
and for $I_{{\bf P}//{\bf \hat{z}}}$ do not simply match and a magnetic signal 
occurs also for ${\bf P}//{\bf \hat{q}}_{\perp}$, in contrast to what is expected 
in Eq.~\ref{master-eq2bis}-\ref{master-eq3}. This discrepancy could be related to 
the fact that the CC picture of Figs.~\ref{fig-CCpatterns}.c-d assumes perfect 
$\rm CuO_2$ planes and, for instance, does not incorporate the dimpling of CuO$_2$ 
planes in the $\rm YBa_2Cu_3O_{6+x}$ system. Further, spin degrees of freedom 
might also play a role in producing in-plane magnetic moments (necessary to account 
for a non-zero signal for ${\bf P}//{\bf \hat{q}}_{\perp}$), by spin-orbit 
scattering as proposed in Ref.~\cite {Vivek06} for the CC-phase.

Considering the reported neutron data \cite{Fauque06}, the observed magnetic order is qualitatively 
consistent with the existence of a circulating current state proposed by C.M. Varma to
account for the pseudo-gap \cite{Varma97,Varma99,Simon02,Varma06}. The neutron data 
nevertheless suggest that a more realistic description of the $\rm CuO_2$ should be 
included, in addition to spin degrees of freedom. It is also important to emphasize that, 
if, on the one hand, the neutron data support the CC-phase model, on the other hand,
alternative interpretations cannot be ruled out considering the limited amount of data 
available.

At that stage, one could consider, as an example, a ferromagnetic order with magnetic  
Cu moment aligned perpendicular to the $\rm CuO_2$ planes. The magnetic intensity at 
${\bf q}$=(002) should vanish as well, whereas a signal should be observed at the Bragg 
reflection ${\bf q}$=(01L), as observed experimentally. Furthermore, the L-dependence 
reported in Fig.~\ref{figUD61}.e is dominated by the cosine square term, corresponding  
to "ferromagnetically" coupled planes. Since the studied Bragg reflections are at 
rather small $|q|$, one cannot distinguish between an evolution at larger vector 
controlled by a term $\sim \frac{1}{q^2}$ for circulating currents or by a term 
$\sim (1-\frac{L^2}{q^2}) F^2_{Cu}(q)$ for ferromagnetic spins (for instance). 
However, this cannot account for the magnetic intensity observed 
for ${\bf P}//{\bf \hat{q}}_{\perp}$ (Fig.\ref{figUD61}.a-c). Another spin model could be 
a magnetic order involving staggered spins on oxygen sites which could also account for the 
observed magnetic scattering (see Ref.~\cite{Fauque06}). Thus,  a spin order cannot be 
ruled out on pure experimental grounds. 
The alternative spin orders have nevertheless two main drawbacks: (i) they should have 
typically been detected by local probe measurements such as NMR or $\mu SR$, (ii) no theoretical 
model has been developed, so far, suggesting this could be relevant for cuprates.

It is clear that more work is needed to determine the exact magnetic 
structure factor and the evolution of the magnetic intensity  at larger wave vector. 
Further experiments will require a study of the magnetic signal with different   
experimental set-up: (i)  by using also a polarizing mirror to analyze the 
final polarization (ii) by performing the 3D polarization analysis\cite{cryopad} 
(using CRYOPAD for instance). Another advantage of CRYOPAD is the suppression of 
the magnetic guide field at the sample position which allows to study the magnetic 
response in the superconducting state. Further, an external magnetic field could 
be applied since it should couple to spins, but should have a minor effect 
(if any) on the circulating currents. 

As a conclusion, we have seen that the neutron polarized scattering is a unique technique 
that has been used to search for the existence of circulating currents in cuprates yielding 
tiny magnetic field distribution to which neutron spin can couple. While the search for flux 
phase and DDW phase remains unsuccessful, there are now experimental evidences
of a magnetic order hidden in the pseudo-gap state of high-$\rm T_c$ superconducting 
cuprates, supporting the circulating current state proposed by C.M. Varma in his theory of 
the pseudo-gap phase.


\end{document}